\documentclass[11pt]{article}
\usepackage[T1]{fontenc}
\usepackage[utf8]{inputenc}
\usepackage[margin=0.9in]{geometry}

\usepackage[margin=0.9in]{geometry}
\usepackage{listings, fancyvrb, multicol}
\usepackage{amsmath, amsthm}
\usepackage{url}
\usepackage{hyperref}
\usepackage{enumitem}
\usepackage{times}
\usepackage{color,xcolor}
\usepackage{authblk}
\usepackage{algorithm}
\usepackage[noend]{algpseudocode}
\usepackage{graphicx}
\usepackage{mathtools}

\algrenewcommand\algorithmicindent{1em}

\algnewcommand{\algcomment}[1]{\qquad{\color{blue}\emph{// #1}}}    
\algnewcommand{\alglinecomment}[1]{{\color{blue}\emph{// #1}}}      

\algnewcommand\alglocal{\textbf{local }}                            
\algnewcommand\algreturn{\textbf{return }}                          
\algnewcommand\algeach{\textbf{each }}                              
\algnewcommand\algretire{\textbf{retire process}}                   

\algnewcommand\algmodassign{\textbf{write}}                         
\algnewcommand\algwritemod[2]{\algmodassign(#1, #2)}              
\algnewcommand\algto{\textbf{to }}                  
\algnewcommand\algis{\textbf{is}}                                   
\algnewcommand\algnot{\textbf{not}}                                 
\algnewcommand\algin{\textbf{in}}                                   
\algnewcommand\algempty{\textbf{empty}}                             

\algnewcommand\algand{\textbf{ and }}                               
\algnewcommand\algor{\textbf{ or }}                                 
\algnewcommand\algassign{\ensuremath{\gets} }                       

\algnewcommand\algtrue{\textbf{true}}                               
\algnewcommand\algfalse{\textbf{false}}                             
\algnewcommand\algnull{\ensuremath{\perp}}                          

\algnewcommand\algarray[1]{\textnormal{array}\ensuremath{\langle}#1\ensuremath{\rangle}}    
\algnewcommand\algmod[1]{\textbf{mod}\ensuremath{\langle}#1\ensuremath{\rangle}}            

\algnewcommand\algorithmicwith{\textbf{with}}
\algnewcommand\algorithmicread{\textbf{read}}
\algnewcommand\algorithmicas{\textbf{as}}
\algnewcommand\Read[2]{\State \alglocal #2 \algassign \algorithmicread(#1)}
\algnewcommand\EndRead{}

\algnewcommand\algorithmicinparallel{\textbf{in parallel}}
\algblockdefx[PARFOR]{ParallelFor}{EndParallelFor}[1]
{\algorithmicfor\ #1\ \algorithmicdo\ \algorithmicinparallel}
{\algorithmicend\ \algorithmicfor}

\makeatletter
\ifthenelse{\equal{\ALG@noend}{t}}%
{\algtext*{EndParallelFor}}
{}%
\makeatother

\mathtoolsset{showonlyrefs}

\newcommand{\multipram}[0]{\emph{multiprefix} CRCW \ensuremath{\mathsf{PRAM}}\renewcommand{\multipram}[0]{multiprefix CRCW \ensuremath{\mathsf{PRAM}}}}

\definecolor{ao}{rgb}{0.0, 0.5, 0.0}

\newcommand{\SC}{\texttt{SC}}
\newcommand{\LL}{\texttt{LL}}
\newcommand{\VL}{\texttt{VL}}

\newcommand{\localpush}{\texttt{local\_push}}
\newcommand{\localpop}{\texttt{local\_pop}}
\newcommand{\push}{\texttt{push}}
\newcommand{\pop}{\texttt{pop}}

\newcommand{\allocate}{\texttt{allocate}}
\newcommand{\alloc}{\texttt{allocate}}
\newcommand{\myfree}{\texttt{free}}
\newcommand{\allocprivate}{\texttt{allocate\_private}}
\newcommand{\myfreeprivate}{\texttt{free\_private}}

\newcommand{\hide}[1]{}
\newcommand{\var}[1]{\texttt{#1}}

\newcommand{\cfont}[1]{\mbox{\tt\bf\small #1}}

\makeatletter
\lst@Key{countblanklines}{true}[t]%
{\lstKV@SetIf{#1}\lst@ifcountblanklines}

\lst@AddToHook{OnEmptyLine}{%
	\lst@ifnumberblanklines\else%
	\lst@ifcountblanklines\else%
	\advance\c@lstnumber-\@ne\relax%
	\fi%
	\fi}
\makeatother

\begin{document}

\title{Concurrent Fixed-Size Allocation and Free in Constant Time}

  \author[1]{Guy E. Blelloch}
  \author[1]{Yuanhao Wei}

  \affil[1]{Department of Computer Science, Carnegie Mellon University, USA}

  \affil[ ]{\textit{\{guyb,yuanhao1\}@cs.cmu.edu}}

  \date{}

\maketitle
  
  \begin{abstract}
Our goal is to efficiently solve the dynamic memory allocation problem in a concurrent setting where processes run asynchronously.  
On $p$ processes, we can support allocation and free for \emph{fixed-sized} blocks with $O(1)$ worst-case time per operation, $\Theta(p^2)$ \emph{additive} space overhead, and using only single-word read, write, and CAS.
While many algorithms rely on having constant-time fixed-size \alloc{} and \myfree{}, 
we present the first implementation of these two operations that is constant time with reasonable space overhead.
\end{abstract}

  
  
  
  \newcounter{results}
\newcommand{\result}[2]{\refstepcounter{results} \vspace{.03in}{\flushleft\textbf{Result \theresults~(#1)}: \emph{#2}}
\vspace{.05in}}

\section{Introduction}

Dynamic memory allocation is a very important problem that plays a role in many data structures. Although some data structures require
variable sized memory blocks, many are built on fixed sized blocks,
including linked lists and trees.  

Our goal is to efficiently solve the dynamic memory allocation problem for
fixed sized memory blocks in a concurrent setting.  
An \alloc{}() returns a reference to a block, and a
\myfree{}$(p)$ takes a reference $p$ to an block.  
We say a block is \emph{live} at some point in the execution
history if the last operation that used it (either a \myfree{}
that took it, or \alloc{} that returned it) was an allocate.
Otherwise the block is \emph{available}.  As would be expected, an
\alloc{} must transition the returned block from available to
live (i.e., the operation cannot return a block that is already live),
and a \myfree{}$(p)$ must be applied to a live block $p$ (i.e., the
operation cannot free a block that is already available).
In our setting, processes run asynchronously and may be delayed arbitrarily.

We describe a concurrent linearizable implementation of the fixed-sized
memory allocation problem with the following properties.

\result{Fixed-sized Allocate/Free}{
\label{result:alloc}
On $p$ processes, we can support linearizable \cfont{allocate} and
\cfont{free} for fixed sized blocks of $k \geq 2$ words from a memory pool
consisting of $m$ blocks, with
  \begin{enumerate}
   \item references that are just pointers to the block (i.e., memory addresses), \label{item:allocref}
   \item $O(1)$ worst-case time for each operation, \label{item:alloctime}
   \item a maximum of $m - \Theta(p^2)$ live blocks, \label{item:live}
   \item $\Theta(p^2)$ extra space for internal metadata, \label{item:allocspace}
   \item single-word (at least pointer-width) read, write, and CAS. \label{item:allocptr}
  \end{enumerate}
}

Our notion of time/space complexity includes both local and shared instructions/objects.
Limiting ourselves to pointer-width atomic instructions means that we 
do not use unbounded sequence numbers or hide bits in pointers.
Achieving the last property of Result \ref{result:alloc} require using a very recent result on implementing LL/SC from pointer-width CAS objects.

We are not addressing the problem of supporting an unbounded
number of allocated cells.  Indeed this would seem to be impossible without
some assumption of how the operating system gives out large chunks of
memory to allocate from. 
Given a memory pool with $m$ blocks, our algorithm is able to allocate $m-\Theta(p^2)$ live blocks before needing to request more memory from the operating system.
In our algorithm, the memory allocator could potentially read from a memory block that
is live, but it works correctly regardless of what the user writes into the memory blocks.
We believe these kinds of accesses are reasonable.

Intuitively, \emph{lock-freedom} means that as long as processes continue executing operations, some process will make progress. 
\emph{Wait-freedom} means that each process will make progress on its operation after a finite number of its own steps.
Our algorithm takes $O(1)$ time for \alloc{} and \myfree{} which implies wait-freedom.

While there has been work on lock-free memory allocation~\cite{michael2004scalable, gidenstam2010nbmalloc, seo2011sfmalloc}, the only work that we know of that is wait-free is by Aghazadeh and Woelfel~\cite{aghazadeh2016tag}.
Aghazadeh and Woelfel's \texttt{GetFree} and \texttt{Release} operations can be used to implement \alloc{} and \myfree{} in constant time, but their algorithm requires $\Omega(mp^3)$ space, making it inpractical in a lot of applications.
We guarantee the same time complexity and wait-freedom while using significantly less space.
While our algorithm is mostly a theoretical contribution, the constants are small and we believe it will be fast in practice as well.

This work complements the large body of work on concurrent memory reclamation (\cite{mckenney2008rcu, ramalhete2017brief, fraser2004practical, herlihy2002rop:seejournalversion, brown2015reclaim, michael2004hazard} just to cite a few). The memory reclamation problem is about determining \emph{when} a block is safe to free in a concurrent setting whereas the memory allocation problem is about \emph{how} to allocate and free memory blocks.
There has been recent progress in designing memory reclamation algorithms that are wait-free and ensure bounded space~\cite{Sundell05, blelloch2020concurrent, nikolaev2020universal, lee2010fast, plyukhin2015, aghazadeh2016tag}.
With wait-free memory reclamation, the memory allocation part becomes the new limiting factor, which motivates the need for wait-free dynamic memory allocation. 
Applications of our memory allocation result include wait-free universal constructions (\cite{fatourou2011highly, herlihy1993methodology} to name a few), wait-free queues~\cite{ramalhete2017poster, kogan2011queue} and wait-free stacks~\cite{goel2016wait}. 

Our algorithm is fairly simple.
It maintains a private pool of memory blocks for each process as well as a shared pool.
In the common case, most calls to \alloc{} and \myfree{} will be handled directly by the private pools.
This technique is widely used and has been shown to be fast in practice in both lock-free~\cite{michael2004scalable, gidenstam2010nbmalloc, seo2011sfmalloc} and lock-based allocators~\cite{berger2000hoard, bonwick2001magazines}.
Blocks are transferred between shared and private pools occasionally to make sure there are not too many or too few free blocks in each private pool.
Having too many blocks in a private pool weakens the bound in Item \ref{item:live} of Result \ref{result:alloc} because these blocks cannot be allocated by other processes.
To reduce the amount of accesses to the shared pool, blocks are transferred between the shared and private pools in the granularity of \emph{batches}, each containing $\ell \in \Theta(p)$ blocks.
Our shared pool is implemented using a variant of the P-SIM stack~\cite{fatourou2011highly} that we modified to include garbage collection.
One challenge when using the P-SIM stack to implement the shared pool is that the P-SIM stack itself requires dynamic memory allocation, which seems to make the problem recursive.
We observe that it is safe for the data structures in the shared pool to allocate memory from the same private pools as the user.
Special care is needed to ensure that the private pools always have enough blocks to service both the user and the shared pool.
While our memory allocator is mostly a theoretical contribution, the constants are small and we believe it will be fast in practice as well.

Section \ref{sec:model} describes our concurrency model and defines some key terms.
In Section \ref{sec:related}, we compare our algorithm with existing work on lock-free memory allocation and give a high level description of the P-SIM stack.
Our algorithm for Result \ref{result:alloc} is presented in Section \ref{sec:alloc} and we conclude in Section \ref{sec:conclusion}.

\section{Model}
\label{sec:model}


A \emph{Compare-and-Swap} (CAS) object stores a value that can be accessed through two operations, read and CAS.
The read operation returns the value that is currently stored in the CAS object.
The CAS operation takes a new value and an expected value and atomically writes the new value into the CAS object if the value that is currently there is equal to the expected value.
The CAS operation returns true if the write happens and false otherwise.
We say a CAS operation is \emph{successful} if it returns true and \emph{unsuccessful} otherwise.

A \emph{Load-Linked/Store-Conditional} (LL/SC) object stores a value and supports three operations: LL, VL (validate-link), and SC.
Let $O$ be an LL/SC object.
The LL operation simply returns the current value of the LL/SC object.
An SC operation by process $p_i$ on $O$ takes a new value and writes it into $O$ if there has not been a successful SC operation on $O$ since the last LL by process $p_i$.
We say that the SC is \emph{successful} if it writes a value into $O$ and \emph{unsuccessful} otherwise.
Successful SC operations return true and unsuccessful ones return false.
A VL operation by process $p_i$ returns true if there has not been a successful SC operation on $O$ since the last LL by process $p_i$.

We work in the standard asynchronous shared memory model~\cite{attiya2004distributed} with $P$ processes communicating through base objects that are either registers, CAS objects, or LL/SC objects. Processes may fail by crashing.
All base objects are word-sized and we assume they are large enough to store pointers into memory.
We do not hide any extra bits in pointers.

We also use the standard notions of \emph{history}, \emph{linearizability}, and \emph{wait-freedom}. An \emph{history} is a sequence of invocation and responses of high-level operations. 
We say that an operation in a \emph{history} is completed if its invocation has a matching response.
The \emph{execution interval} of a completed operation in a history is the interval between its invocation and response.
For non-completed operation, its execution interval is from its invocation to the end of the history.
A history is considered \emph{linearizable}~\cite{herlihy1990linearizability} if each completed operation appears to take affect atomically at some point in its execution interval.
Each non-completed operations either takes effect atomically in its execution interval or not at all.
A set of operations is said to be \emph{linearizable} if every history consisting of these operations is linearizable.
All implementations that we discuss will be \emph{wait-free}. This means that each operation by a non-faulty process $p_i$ is guaranteed to complete within a finite number of instructions by $p_i$. 

The \emph{time complexity} of an operation $O$ is the maximum number of instructions (including both shared memory and local instructions), over all possible executions, that a process takes to complete an instance of $O$.
The \emph{space complexity} of an implementation is the total number of word-sized registers, CAS objects, LL/SC objects and local variables that it uses.

  
\section{Related Work}
\label{sec:related}

\subsection{Lock-free Memory Allocation}

Many existing lock-free memory allocators~\cite{michael2004scalable, gidenstam2010nbmalloc, seo2011sfmalloc} are based on a well-known lock-based allocator called Hoard~\cite{berger2000hoard}.
To handle variable size memory allocation, these allocators instantiate multiple instances of their algorithm, one for each size class.
Each instance can be thought of as a fixed size memory allocator.
At a high level, each instance maintains a global heap as well as a private heap for each process which is similar to our shared and private pools.
To reduce the amount of accesses to the global heap, blocks are transferred between the global and private heaps in large chunks called \emph{superblocks}.
Superblocks are contiguous pieces of memory which usually spans some multiple of the system's page size.
Superblocks are different from our notion of batches because the blocks in a batch are not necessarily contiguous.
Another difference is that in these existing algorithms~\cite{michael2004scalable, gidenstam2010nbmalloc, berger2000hoard}, an additional word is added to each memory block to store a pointer back to the superblock from which it was allocated.
When a memory block is passed to \myfree{}, it gets added back to its original superblock, which could belong to the private heap of another process.
In our algorithm, processes free memory back to their own private pool.
When all the blocks from a superblock have been freed, the superblock gets freed back to the OS so that it is available to be allocated for a different size class.
When a process is unable to find a free block in its private heap and in the global heap, the process allocates a new superblock from the OS.

To measure memory overhead, Berger et al.~\cite{berger2000hoard} introduces the notion of \emph{memory blowup}.
Memory blowup is defined to be the difference between the maximum amount of memory required by a concurrent allocator and the maximum amount of memory required by an ideal sequential allocator on the same execution trace.
For the fixed size allocation problem, these allocators~\cite{michael2004scalable, gidenstam2010nbmalloc, berger2000hoard} incur constant factor memory blowup as well as $\Theta(pS)$ additive memory blowup, where $S$ is the size of their superblocks.
Using this terminology, our algorithm incurs $\Theta(p^2)$ additive memory blowup.

SFMalloc~\cite{seo2011sfmalloc} introduces many practical optimizations for improving cache performance and reducing synchronized instructions.
It is lock-free and leverages ideas from Hoard as well as several other papers.
LFMalloc~\cite{dice2002mostly} is a partly lock-free allocator that requires special operating system support.

None of these existing allocators support \alloc{} and \myfree{} in constant time even in the fixed-size blocks setting.

\subsection{Shared Pool}

The shared pool in our memory allocator can be implemented using any dynamic-sized concurrent bag data structure.
We chose to start with the P-SIM stack because it naturally satsifies most of the properties we are aiming for.
Some alternatives include Kogan and Petrank's wait-free queue~\cite{kogan2011queue}, Ramalhete and Correia's wait-free queue~\cite{ramalhete2017poster} and Goel, Aggarwal and Sarangi's wait-free stack~\cite{goel2016wait}.
The first alternative requires unbounded sequence numbers and it is not clear how to get rid of them, and the latter two require $\Omega(p)$ time per operation.
Overall, the P-SIM stack required the least modifications to achieve the properties we want.

\subsection{P-SIM Stack}

In this section, we give a high-level description of the P-SIM stack which will be helpful for understanding Section \ref{sec:shared}.
P-SIM is a universal construction, which is a method of transforming any sequential data structure into a concurrent data structure.
The P-SIM stack uses P-SIM to turn a standard linked-list based sequential stack into a concurrent stack.
P-SIM works by maintaining a state record \var{S} which stores the current state of the shared object as well as other information.
Processes apply changes locally and then update this state record using \LL{}/\SC{}.
To run an operation on the P-SIM stack, the process first announces the operation it wishes to perform (either a \push{} or a \pop{}) and then reads the current state record using \LL{}(\var{S}).
Next, it checks which processes have announced an operation and for each operation that is announced and not yet applied, it applies the operation locally to compute a new local state record \var{ls}.
Finally, the process tries to write \var{ls} into \var{S} with an \SC{}.
If this \SC{} succeeds, then its operation has been applied.
Otherwise, it tries again starting from the \LL{} step.
If the \SC{} fails a second time, then its operation is guaranteed to have been applied by another process.
This high-level description leaves out many important details, but we believe it captures the essential information needed to understand Section \ref{sec:shared}.
For the full algorithm, we refer the reader to the original paper~\cite{fatourou2014highlyfull}.








  \section{Wait-free Fixed-sized Memory Allocation}
\label{sec:alloc}

For some constant $l \in \Theta(p)$, our data structure consists of local private pools that each hold $\Theta(l)$ blocks and a shared pool that maintains a stack of 
\emph{batches}, each containing $l$ blocks.
Pushes and pops by any process from the shared stack take $O(p)$ time.  To amortize their cost, they can be broken into $p$ steps of $O(1)$ time each.

In Section \ref{sec:shared}, we implement the shared stack by starting with Fatourou and Kallimanis's P-SIM stack~\cite{fatourou2014highlyfull} and adding memory management.
In Section \ref{sec:private}, we describe how to implement the private pools.


\subsection{Shared Pool}
\label{sec:shared}

The shared pool is maintained as a shared stack where each node stores a pointer to a batch.
Using a modified version of Algorithms 2 and 3 from~\cite{fatourou2014highlyfull}, we can implement a stack of pointers with the following properties:

\result{Shared Stack}{
\label{result:stack}
Given a concurrent allocator satisfying Result \ref{result:alloc} with parameter $k \geq 2$, assuming that reading from a memory block that has already been freed returns an arbitrary value, on $p$ processes, we can support linearizable \cfont{push} and \cfont{pop} with
  \begin{enumerate}
   \item $O(p)$ worst-case time for each operation, \label{item:time}
   \item at most $2p$ calls to \cfont{allocate} and $2p$ calls to \cfont{free} in each operation, \label{item:2p-alloc}
   \item $Mk + \Theta(p^2k)$ space (where $M$ is the number of nodes in the stack), \label{item:space}
   \item single-word (at least pointer-width) read, write, and CAS, \label{item:atomics}
  \end{enumerate}
}

At first glance, this result may seem circular because we use Results \ref{result:alloc} and \ref{result:stack} to implement each other.
However, this is the key trick in our algorithm.
We explain why this works in more detail in Section \ref{sec:private}, but the high level idea is that the data structures in the shared pool can actually allocate memory from the same private pools as the user.
For this idea to work, Property \ref{item:2p-alloc} of Result \ref{result:stack} is crucial.

In most real-world systems, accessing memory that has already been freed is not allowed.
However, these kinds of accesses are reasonable in our setting because the shared stack is internal to our memory allocator.
Whenever a node is popped off the shared stack and freed, the memory is not freed back to the operating system, instead it is made available to be allocated to the user.

  \begin{figure}
  \begin{lstlisting}[linewidth=\columnwidth, xleftmargin=5.0ex, numbers=left]
struct Request {
  void *func;
  ArgVal arg; };

struc StRec {
  State st;
  boolean applied[1..p];
  RetVal rvals[1..p]; };

shared Integer Toggles = 0;
shared StRec Pool[1..p+1][1..2];
shared StRec* S = &Pool[p+1][1];

thread_local Integer toggle = @$2^i$@;
thread_local Integer index = 0;

RetVal PSimApplyOp(Request req, ThreadId i) {
  Announce[i] = req;
  Add(Toggles, toggle);
  toggle = -toggle;
  Attempt(i);
  return S.rvals[i]; }

void Attempt(ThreadId i) {
  bool ltoggles[1..p];                  @\label{line:ltoggles}@
  StRec *ls_ptr;                        @\label{line:lsptr}@
  for j = 1 to 2 do {                   @\label{line:attemptloop}@
    ls_ptr = LL(S);                     @\label{line:ll}@
    Pool[i][index] = *ls_ptr;
    if(VL(S) == 0)
      continue;
    ltoggles = Toggles;
    for a = 1 to p do {                                     @\label{line:help-loop}@
      // If @$p_i$@ has a request that has not been applied yet
      if(ltoggles[a] != Pool[i][index].applied[a]) {
        // Apply the request that has not been applied yet
        apply Announce[a] on Pool[i][index].st                    @\label{line:applyop}@
        and store the return value into Pool[i][index].rvals[a]; }
      Pool[i][index].applied[a] = ltoggles[a]; }
    if(SC(S, & Pool[i][index]))               @\label{line:sc}@
      index = (index mod 2) + 1; } } }
\end{lstlisting}
  \caption{P-SIM Universal Construction (Algorithm 2 from \cite{fatourou2014highlyfull})}
  \label{fig:psimstack}
\end{figure}

  \begin{figure}[!t!h]
  \begin{lstlisting}[linewidth=\columnwidth, xleftmargin=5.0ex, numbers=left,firstnumber=42]
struct Node { Data data; Node *next; };

struct State { Node *top };

void Push(ArgVal arg, ThreadId i) {
  PSimApplyOp(<local_push, arg>, i); }

Node* Pop(ThreadId i) {
  return PSimApplyOp(<local_pop, $\bot$>, i); }

void local_push(State* pst, ArgVal arg) {    @\label{line:localpush}@
  nd = allocate a @\texttt{new}@ node;                  @\label{line:alloc}@
  nd->data = arg;
  nd->next = pst->top;
  pst->top = nd; }

Node* local_pop(State* pst) {                @\label{line:localpop}@
  Node* ret = pst->top;
  if(pst->top != @$\bot$@)
    pst->top = pst->top->next;               @\label{line:read-next}@
  return ret; }
\end{lstlisting}
  \caption{P-SIM Stack (Algorithm 3 from \cite{fatourou2014highlyfull})}
  \label{fig:psimstack2}
\end{figure}

The pseudo-code from~\cite{fatourou2014highlyfull} is reproduced in Figures \ref{fig:psimstack} and \ref{fig:psimstack2} for ease of reference.
A high-level description of how the P-SIM stack works was given in Section \ref{sec:related}.

By default, the P-SIM stack satisfies Properties \ref{item:time} and \ref{item:2p-alloc} of Result \ref{result:stack}, but it does not fulfill Properties \ref{item:space} and \ref{item:atomics} because it never frees nodes, and uses both LL/SC and fetch-and-add. 
Fetch-and-Add is only used for a practical optimization and the array \var{toggles} can instead be implemented as an array of registers without affecting any theoretical bounds.
In~\cite{fatourou2014highlyfull}, the pointer-width LL/SC object $S$ is simulated with a timestamped CAS object.
Since we want to avoid unbounded sequence numbers, we instead use the LL/SC from CAS algorithm from Result 1 of~\cite{atomiccopy}.
This algorithm implements LL, VL, and SC in constant time, requires only pointer-width read, write, and CAS, and uses $O(cp^2)$ space where $c$ is the number of outstanding LL/SC operations per process.
In our case, $c = 1$ so it uses only $O(p^2)$ space.
There has been a lot of work on implementing LL/SC from CAS, but~\cite{atomiccopy} is the only one we are aware of that maintains the other properties in Result \ref{result:stack}.
If word size is large enough to store $(4\log{} p)$ bits, then we could also have used the LL/SC from CAS algorithm from~\cite{jayanti2003efficient}.
After these modifications, the only property that the P-SIM stack does not satisfy is Property \ref{item:space}.

Adding memory management to the P-SIM stack is slightly more involved.
Note that all nodes are allocated on line \ref{line:alloc} of \localpush{}.
In this paragraph, we show how to ensure that there are at most $M+O(p^2)$ allocated nodes that have not been freed, where $M$ is the size of the stack.
To begin, note that all calls to \localpush{} and \localpop{} happen between the \LL{} on Line \ref{line:ll} and the \SC{} on Line \ref{line:sc}.
These local operations ``take effect'' only if the \SC{} succeeds.
Between the \LL{} and the \SC{}, we modify the algorithm so that it keeps track of the nodes that it locally pushed and the nodes that it locally popped.
If the \SC{} fails, then it frees all the nodes that were locally pushed because they will no longer appear in the global state, and if the \SC{} succeeds, then it frees all the nodes that were locally popped because they are now also popped from the global state.
These changes on their own are not enough for correctness because after a process performs a successful \SC{} and frees a node, it is possible for the contents of that node to be accessed by a process working on an outdated copy of the global state.
The contents of a node are only accessed on line \ref{line:read-next} at the read of \var{pst->top->next}.
Even though we allow the algorithm to read from memory that it has freed, the access on line \ref{line:read-next} is still dangerous because it could lead to an invalid pointer being dereferenced in future \localpop{} operations.
We prevent this by performing an \VL(S) immediately after line \ref{line:read-next}.
If this \VL{} is successful, then the node that was previously accessed has not yet been freed, so the read is safe.
Otherwise, the \SC{} on line \ref{line:sc} is guaranteed to fail, so we can jump to the next iteration of the loop on line \ref{line:attemptloop}.
If we fail on a \VL{}, we also have to free all the nodes that were allocated by \localpush{} operations.


With these changes, it is straightforward to verify that all the properties in Result \ref{result:stack} are satisfied.
To show the space complexity in Property \ref{item:space}, we first argue that there are at most $M + O(p^2)$ nodes that have been allocated and not freed, where $M$ is the number of nodes reachable from the top of the shared stack (i.e. the node pointed to by \var{S->st.top}).
Each process has at most $p$ nodes that it allocated, which have not been freed and are not part of the shared stack.
Furthermore, each process has at most $p$ nodes that it popped from the shared stack, but has not yet freed.
Therefore, there are at most $O(p^2)$ nodes that are not part of the shared stack, so the overall number of nodes is at most $M + O(p^2)$.
Each node takes up $k$ words due to the granularity of \allocate{}.
$O(p^2)$ space is needed for the other variables in the universal construction, so the overall space usage is $Mk + O(p^2k)$.

\subsection{Private Pools}
\label{sec:private}

We manage the private pools as follows.  Each pool has between zero and two full batches with $\ell$ blocks each.  It also has a current working batch that is partially filled and contains between $1$ and $\ell-1$ elements, and possibly one of either a popped batch or a pushed batch from the shared pool that is currently being processed. 
Due to the deamortization, operations on the shared pool will complete $p$ \alloc{} or \myfree{} operations after they start.
We maintain a \var{num\_batches} variable that represents the number of full batches plus one if there is a pop currently being processed (started but not completed).  This variable always has a value of 1 or 2.  The pseudocode for \alloc{} and \myfree{} is given in Figure~\ref{fig:allocfree}.

At a high level, each process performs \alloc{} by removing a block from the process's private pool and \myfree{} by adding a block to the process's private pool.
The shared pool is used to ensure that the private pools do not get too big or too small.
We implement the shared pool using the shared stack from Result \ref{result:stack}.
Since \push{} and \pop{} operations on the shared stack take $O(p)$ time, they can be broken up into $p$ steps, each of which take constant time.
Each \alloc{} or \myfree{} by process $p_i$ performs one step towards any ongoing \push{} or \pop{} operations from the shared pool by $p_i$.
We will show later that there is at most one such ongoing \push{} or \pop{} per process.

Since the shared stack requires constant time memory allocation, we implement a simplified version of \alloc{} and \myfree{} called \allocprivate{} and \myfreeprivate{}.
This simplified version will only be called by \push{} or \pop{} operations on the shared stack.
These two operations are implemented the same way as regular \alloc{} and \myfree{} except with all the shared pool operations removed.
This means \allocprivate{} simply returns a block from the process's private pool and \myfreeprivate{} simply adds the freed block to the process's private pool.
Psuedocode is given in Figure~\ref{fig:allocfree2}.

When a process has only $\ell$ blocks left in its private pool, it begins popping a batch from the shared pool (Line \ref{line:delayedpop}) provided it is not already in the middle of such a \pop{}.
We require $\ell$ to be at least $3p$ because it takes $p$ \alloc{} or \myfree{} operations to \pop{} a batch from the shared pool and this \pop{} operation could call \allocprivate{} up to $2p$ times (Item \ref{item:2p-alloc} of Result \ref{result:stack}).
A process needs to have enough blocks in its private pool to serve all the allocate requests while it waits for the shared \pop{} to complete.
Once the \pop{} completes, the process gains $\ell$ new blocks for its private pool.

When a process has $3\ell$ total blocks in its private pool, it begins pushing a batch to the shared pool.
Now we show that there is at most one ongoing delayed \push{} or \pop{} per process.
When a delayed \push{} begins (Line \ref{line:delayedpush}), we consider the batch being pushed to no longer be part of the process's private pool.
The delayed \push{} completes within $p$ calls to \alloc{} or \myfree{} and calls \myfreeprivate{} at most $2p$ times, so the private pool's size won't increase back to $3\ell$ until after the delayed \push{} completes.
By the same argument, after a delayed \pop{} is invoked, the private pool's size won't increase to $3\ell$ until after the delayed \push{} completes, and after a delayed \push{} is invoked, the private pool's size won't decrease to $\ell$ until after the \push{} completes.
Therefore, each process has at most one ongoing delayed \push{} or \pop{}.


By managing the private pools in the manner, we ensure that each process's private pool always has between $1$ and $3\ell$ blocks, so it will never run out of blocks to allocate and will never hold on to too many blocks.

Finally, we explain the details of how batches are implemented.
Each batch is maintained as a stack of blocks.
Each block stores a pointer to the next block in the batch and the batch is represented by a pointer to the first block in the batch.
Borrowing memory from the blocks allows us to avoid allocating memory on \push{} operations. 
Similarly, we implement the stack of batches (\var{local\_batches}) by borrowing an additional word from the first block of each batch to store the next pointer for the \var{local\_batches} stack.
This implementation works because each block appears in at most one batch and each batch appears in at most one process's \var{local\_batches} stack.
Each block needs to have enough space for two pointers which is another reason why we require $k \geq 2$.

It is easy to verify that this algorithm satisfies Properties \ref{item:allocref}, \ref{item:alloctime} and \ref{item:allocptr} of Result \ref{result:alloc}.
For Property \ref{item:allocspace}, the extra space is needed to implement the shared stack in Result \ref{result:stack} and some of the local variables in Figure \ref{fig:allocfree}.
For Property \ref{item:live}, we know that the private pools contain at most $9p^2$ blocks in total which means that as long as there are at least $9p^2$ blocks that have been freed and not allocated, the shared stack cannot be empty.




  \begin{figure}
  \begin{lstlisting}[linewidth=\columnwidth, xleftmargin=5.0ex, numbers=left]
using batch = stack<block>;
shared_stack<batch> shared_pool;
thread_local stack<batch> local_batches; 
thread_local stack<block> current_batch;
int num_batches = 2;

block* allocate() {
  if (current_batch.is_empty()) {
    current_batch = local_batches.pop();
    if (num_batches == 1) 
      delayed local_batches.push(shared_pool.pop());  @\label{line:delayedpop}@
    else num_batches--; }
  run_delayed_step();
  return current_batch.pop(); }

void free(block* b) {
  if (current_batch.full()) {
    if (num_batches == 2) 
      delayed shared_pool.push(current_batch);       @\label{line:delayedpush}@
    else {
      num_batches++;
      local_batches.push(current_batch); }
    current_batch.clear(); }
  run_delayed_step();
  current_batch.push(b); }            @\label{line:push-block}@
\end{lstlisting}
  \caption{Constant Time Allocate and Free}
  \label{fig:allocfree}
\end{figure}

  \begin{figure}
  \begin{lstlisting}[linewidth=\columnwidth, xleftmargin=5.0ex, numbers=left]
block* allocate_private() {
  if (current_batch.is_empty()) {
    current_batch = local_batches.pop();
    num_batches--; }
  return current_batch.pop(); }

void free_private(block* b) {
  if (current_batch.full()) {
    num_batches++;
    local_batches.push(current_batch);
    current_batch.clear(); }
  current_batch.push(b); }            @\label{line:push-block}@
\end{lstlisting}
  \caption{Allocate and Free used internally by \var{shared\_pool}}
  \label{fig:allocfree2}
\end{figure}

\section{Conclusion}
\label{sec:conclusion}

We presented a constant time algorithm for fixed-size dynamic memory allocation with $\Theta(p^2)$ \emph{additive} space overhead using only single-word read, write, and CAS.
An interesting open problem would be to efficiently support variable-size memory allocation in a wait-free manner.






  \section{Acknowledgments}
  We would like to thank our anonymous reviewers for their helpful comments.

  \bibliography{../../biblio}
 
\end{document}